\documentclass[twocolumn,superscriptaddress,pre,aps]{revtex4}
\usepackage{amsmath,graphicx}

\begin{document}

\title{Numerical Simulation of the Trapping Reaction with Mobile and
  Reacting Traps}
\author{Joshua D. Hellerick}
\affiliation{Department of Physics, Bucknell University, Lewisburg PA
  17837, USA}
\author{Robert C. Rhoades} 
\affiliation{Center for Communications Research, Princeton, NJ 08534, USA}
\author{Benjamin P. Vollmayr-Lee} 
\affiliation{Department of Physics, Bucknell University, Lewisburg PA
  17837, USA}
\date{\today}

\begin{abstract}
  We study a variation of the trapping reaction, $A+B\to A$, in which
  both the traps ($A$) and the particles ($B$) undergo diffusion, and
  the traps upon meeting react according to $A+A\to 0$ or $A$.  This
  two-species reaction-diffusion system is known to exhibit a
  non-trivial decay exponent for the $B$ particles, and recently
  renormalization group methods have predicted an anomalous dimension
  in the $BB$ correlation function.  To test these predictions we
  develop a computer simulation method, motivated by the technique of
  Mehra and Grassberger, that determines the complete probability
  distribution of the $B$ particles for a given realization of the $A$
  particle dynamics, thus providing a significant increase the quality
  of statistics.  Our numerical results indeed reveal the
  anomalous dimension predicted by the renormalization group, and
  compare well quantitatively to precisely known values in cases where
  the problem can be related to a 4-walker problem.
\end{abstract}
\maketitle

\section{Introduction}
\label{sec:intro}

Reaction-diffusion processes with irreversible reactions provide an
important class of far from equilibrium systems.  Interest in these
systems stems from the fact that the particles develop nontrivial
correlations that cannot be described by equilibrium fluctuations, and
these correlations in turn affect the reaction rates and particle
densities.  Applications for these model systems include chemical
reaction kinetics \cite{Rice85}, interface growth models
\cite{Krug91}, aggregation \cite{Spouge88}, domain coarsening
\cite{Bray94}, and population dynamics \cite{Taeuber12}.

In the present work, we consider a two-species process consisting of
the trapping reaction $A+B\to A$, in which $A$ particles, or
``traps,'' catalyze the decay of $B$ particles, and where the traps
additionally react according to $A+A\to 0$ (annihilation) or $A+A\to
A$ (coalescence).  Both particle types $A$ and $B$ undergo diffusion
with corresponding diffusion constants $D_A$ and $D_B$.  This system
has been predicted via renormalization group (RG) methods to exhibit
anomalous dimension in both the $B$ particle density decay
\cite{Howard96,Krishnamurthy03,Rajesh04} and separately in the scaling
of the $BB$ correlation function \cite{VollmayrLee18} for spatial
dimension $d<2$.  The primary focus of this paper is to test these
predictions numerically in one-dimensional systems.  For this purpose
we develop a hybrid Monte Carlo technique that provides the entire $B$
particle distribution for a given realization of the $A$ particles.
This is possible because, as argued below, the $B$ particles remain
locally Poissonian.

For the $A+B\to A$ trapping reaction with mobile but
\textit{non-reacting} traps, the mean-field rate equation predicts the
$B$ particle density to decay exponentially with time. However,
scaling arguments and rigorous bounds confirm that for dimension $d<2$
nontrivial correlations develop between the traps and the surviving
$B$ particles, invalidating the rate equation and causing the $B$
particle density to decay as a stretched exponential $\langle
b\rangle\sim \exp(-\lambda_d t^{d/2})$ with a universal coefficient
$\lambda_d$ \cite{Bramson88,Bray02,Blythe03}.  Here and throughout
angle brackets are used to indicate averages over the
random initial conditions and over the stochastic processes of
reaction and diffusion.

Now consider traps that are additionally reacting according to
\begin{equation}
  A+A\to \begin{cases} A & \text{(coalescence) probability } p \\
    0 & \text{(annihilation) probability } 1-p.
  \end{cases}
  \label{eq:trap_reaction}
\end{equation}
Since the traps are unaffected by the $B$ particles, their dynamics
reduces to the well-studied single-species reaction, where mean-field
rate equations (see below), exact solutions in one spatial dimension
\cite{Torney83,Lushnikov87,Privman97}, and field-theoretic
RG methods \cite{Peliti86,Lee94,Taeuber05} for
general dimension demonstrate that the $A$ particle density decays as
power law (with a multiplicative logarithmic correction in $d=2$).
This decaying trap density then enhances the survival probability of
the $B$ particles, resulting in a power law decay with time, $\langle
b\rangle \sim t^{-\theta}$.  For example, the rate equations, valid
for $d>2$ where diffusion manages to keep the reactants well mixed,
are
\begin{equation}
  \partial_t \langle a\rangle = -\Gamma\langle a\rangle^2,
  \qquad
  \partial_t \langle b\rangle = -\Gamma'\langle a\rangle \langle b\rangle,
\end{equation}
with solutions $\langle a\rangle \sim 1/(\Gamma t)$ and $\langle
b\rangle$ decay exponent determined by the nonuniversal rate
constants, $\theta=\Gamma'/\Gamma$.

For $d< 2$ the depletion caused by reactions competes with diffusion,
developing correlations that modify the reaction rate.  This results
in the trap density decay $\langle a\rangle \sim A_d(D_At)^{-d/2}$
with a universal coefficient $A_d$.  The $B$ particle density in this
fluctuation-dominated case has been studied with Smoluchowski theory
\cite{Krapivsky94}, which is an improved rate equation that
incorporates the depletion with a time-dependent rate constant, and
with RG techniques
\cite{Howard96,Krishnamurthy03,Rajesh04,VollmayrLee18}.  In both cases
the $B$ particle density was found to decay as a power law with a universal
exponent $\theta$ depending only on the diffusion constant ratio
$\delta =D_B/D_A$ and the trap reaction parameter $p$ defined in
Eq.~(\ref{eq:trap_reaction}).  Smoluchowski theory gives
\begin{equation}
  \theta_S = \frac{d}{2-p} \biggl(\frac{1+\delta}{2}\biggr)^{d/2}
  \label{eq:theta_smoluchowski}
\end{equation}
while the RG analysis predicts
\begin{equation}
  \theta = \theta_S + \frac{1}{2} \gamma_b^*
\end{equation}
where $\gamma_b^*$ is an anomalous dimension of order $\epsilon = 2-d$
which stems from a field renormalization of the density
\cite{Krishnamurthy03,VollmayrLee18}.

Recently it was shown by RG methods that an additional anomalous
dimension occurs due to the field renormalization of the $b^2$ density
operator \cite{VollmayrLee18}, with the consequence that the $B$
particle correlation function scales as
\begin{equation}
  C_{BB}(r,t) \equiv
  \frac{\langle b(r,t) b(0,t)\rangle - \langle b(t)\rangle^2}
       {\langle b(t)\rangle^2} \sim t^\phi f(r/\sqrt{t}),
       \label{eq:BBcorrelations}
\end{equation}
where $\phi$ is a universal exponent of order $\epsilon$.  In
contrast, the scaled correlation functions $C_{AA}$ and $C_{AB}$ are
simply functions of $r/\sqrt{t}$ with no time-dependent prefactor.  We
note that $\chi_{BB}(t)\equiv C_{BB}(0,t)$ is a measure of the local
fluctuations, and Eq.~(\ref{eq:BBcorrelations}) predicts that
$\chi_{BB}$ grows as a universal power of time.  In
Ref.~\cite{VollmayrLee18} the exponent $\phi$ was computed to first
order in $\epsilon$.  Additionally, an exact value of $\phi$ was
obtained for the case of $p=\delta=1$ in one spatial dimension by
mapping to a four walker problem \cite{VollmayrLee18} and solving an
eigenvalue problem numerically \cite{Helenbrook18}.

Here we aim to use numerical simulations to test the predicted scaling
form Eq.~(\ref{eq:BBcorrelations}) and to measure the exponents
$\theta$ and $\phi$.  These simulations are challenging since the
window of scaling behavior is limited by transients at early times and
finite size effects and vanishing particle numbers at late times.
In the present work we circumvent the small number statistics of the
$B$ particles by determining the entire $B$ particle probability
distribution conditioned on a particular realization of the $A$
particle dynamics.  Our technique was inspired by and is a converse to
the method of Mehra and Grassberger \cite{Mehra02}, who studied the
trapping reaction by monitoring a single particle and updating the
distribution of traps.
With greatly improved statistical accuracy, we
were able to demonstrate the scaling collapse of the $AA$, $AB$, and
$BB$ correlation functions and measure the dynamical exponents
$\theta$ and $\phi$ to high accuracy.

The layout of this paper is as follows.  In Sec.~\ref{sec:method} we
present our hybrid simulation method, which also serves to define the
model we are considering.  In Sec.~\ref{sec:density} we report our
measurements of the density decay exponent $\theta$ for a variety of
$\delta$ and $p$ values, and compare these to known exact solutions,
RG calculations, and the Smoluchowski approximation. Then in
Sec.~\ref{sec:phi} we present our data for the anomalous dimension
$\phi$, and compare to the RG prediction and the exact solution from
the 4-walker problem, while in Sec.~\ref{sec:correlations} we test the
pair correlation functions for scaling collapse.  Finally, in
Sec.~\ref{sec:summary} we summarize our results and suggest future
work.

\section{Hybrid Monte Carlo and Master Equation Method}

\label{sec:method}

Reaction-diffusion systems are typically simulated via Monte Carlo
methods: a lattice is populated randomly by particles, and then
updated according to the particular rules for reaction and stochastic
diffusion.  Quantities of interest are then averaged over multiple
realizations of the stochastic processes.  Monte Carlo is employed
rather than direct computation of the probabilities in a master
equation because of the impossibility in dealing with such a large
number of configurations.

However, for the trapping reaction the $B$ particles are
non-interacting, and this allows for a much simpler description of the
$B$ particle probabilities.  We use this to construct a hybrid
approach in which we use Monte Carlo for the $A$ particles, but for
each realization of the $A$ particle dynamics we calculate the entire
$B$ particle probability distribution.  This is possible because the
$B$ particle distribution remains Poissonian at each lattice site.

We now define our model for concreteness.  We consider a
$d$-dimensional hypercubic lattice and use a parallel update.  The $A$
and $B$ particles are initially randomly distributed on sites whose
lattice indices sum to an even number.  In a diffusion step each
particle will simultaneously hop in one of the $\pm\hat x_i$
directions along the principle axes of the lattice, so that after an
even (odd) number of steps, the particles reside in the even (odd)
sector of the bi-partite lattice.  Reactions are then performed
subsequent to the diffusion hops.  In the simplest scenario, for any
site containing both $A$ and $B$ particles, the $B$ particles are
removed.  A variant of this rule would be for each $B$ particle to be
removed with probability $p'$.  Any site containing two $A$ particles
reacts according to Eq.~(\ref{eq:trap_reaction}), governed by the
parameter $p$.

When the $A$ and $B$ diffusion constants are equal, both particle
types step simultaneously, resulting in the diffusion constant $D =
\Delta x^2/(2 d \Delta t)$ for a lattice constant $\Delta x$ and a hop
time $\Delta t$.  For unequal diffusion constants we can take an odd
number of multiple steps for one of the species.  For example, if
$\delta = D_B/D_A=3$ we take two steps with the $B$ particles, check
the $A+B\to B$ reaction, take one more step with both particle types,
and then check the reactions again.  For $\delta=2$ we first do the
process just described and then take one more step with both particle
types.  In this way any rational value of the diffusion constant ratio
$\delta$ can be realized.

Our hybrid technique relies on the following two well-known properties
of Poisson distributions:
\begin{itemize}
\item[P1.]  The sum of two independent Poisson distributed random variables
  with mean values $\mu$ and $\nu$ is a Poisson random variate with mean
  $\mu+\nu$.
\item[P2.] The compound of a Poisson distribution with mean $\mu$ and
  a binomial distribution with probability $q$ is a Poisson
  distribution with mean $q\mu$.
\end{itemize}
The second property says that if a number of elements is a Poissonian
random variate and then a random subset of elements are selected with
independent probabilities, the selected number of elements is a
Poissonian random variate.

\begin{figure}
  \includegraphics[width=3.25in]{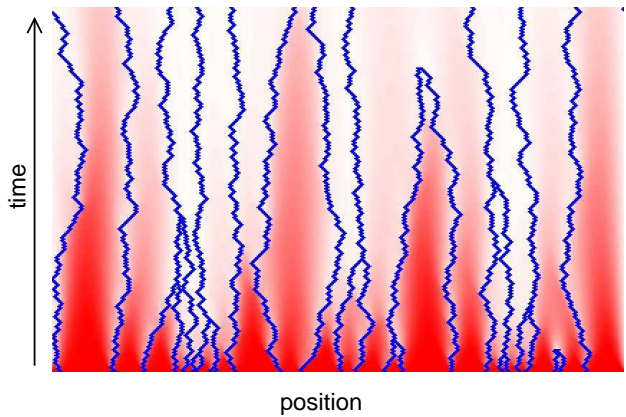}
  \caption{A characteristic segment of our simulation.  The blue lines
    are $A$ particles (traps), which undergo both coalescence and
    annihilation reactions.  The $B$ particle probability distribution
    is shaded in red, with the intensity representing the local
    Poissonian mean.}
  \label{fig:reaction_picture}
\end{figure}

Now consider a Poisson distribution of $B$ particles on site $i$ with
mean value $b_i$.  In the diffusion step the probability of a particle
making the hop to a particular nearest neighbor $j$ is $1/(2d)$.  Thus
from property P2 these particles will contribute a Poissonian
distributed number of particles with mean $b_i/(2d)$ to each of their
neighboring sites.  The new distribution at a particular site $j$ is a
sum of Poisson random variates, thus by property P1 it is Poissonian
with mean given by
\begin{equation}
  b_{j,t+\Delta t} = \frac{1}{2d}\sum_k b_{k,t}
  \label{eq:b_diffusion_update}
\end{equation}
where $k$ are the nearest neighbors of $j$.

To incorporate the trapping reaction, we take
\begin{equation}
  b_{i,t}\to (1-p') b_{i,t}
  \label{eq:b_reaction_update}
\end{equation}
at any site $i$ containing an $A$ particle at time $t$, which derives
from property P2, recalling that each $B$ particle independently
reacts with probability $p'$, or survives with probability $1-p'$.

With this method, an explicit realization of the $A$ particles is
evolved, and simultaneously the local means of the Poissonian $B$
particles are updated by use of Eqs.~(\ref{eq:b_diffusion_update}) and
(\ref{eq:b_reaction_update}).  The computational cost of this method
in comparison to a Monte Carlo simulation of the $B$ particles is the
introduction of a floating point variable that has to be updated at each
lattice site at each time step.  The gain is vastly improved
statistics, particularly for parameter values where $\theta$ is large,
for which the $B$ particle density decays rapidly and Monte Carlo
simulations would yield vanishing particle numbers.

\section{$B$ Particle Density}

\label{sec:density}

We measured the $B$ particle density for one-dimensional
systems with lattice size ranging from $10^6$ up to $3\times 10^7$ sites.
We set $\Delta x = \Delta t = 1$ and used an initial condition of
$\langle a(0)\rangle = 0.5$ for the trap density and without loss of
generality we set $\langle b(0)\rangle$ to unity.

Simulations were performed for diffusion constant ratios
$\delta=D_B/D_A=1/4$, $1/2$, $1$, $2$, and $4$ for both the $A+A\to 0$
($p=0$) and the $A+A\to A$ ($p=1$) trap reactions.  Additionally, for
equal diffusion constants $\delta =1$ we simulated mixed trap
reactions with $p=1/4$, $1/2$, and $3/4$, with $p$ defined in
Eq.~(\ref{eq:trap_reaction}). We also varied the trapping probability
parameter $p'$ in Eq.~(\ref{eq:b_reaction_update}) to confirm the
universality of our results.  The data presented here and below
correspond to $p'=1$.  In each case we performed between 100 and 400
independent runs.  In order for the statistical uncertainties at
different times to be uncorrelated, we used an independent set of runs
for each time value where we collected data.  The onset time for
finite size effects depended strongly on the parameters $\delta$ and
$p$, decreasing with respect to both parameters.  As such, we chose
the system size and simulation run time accordingly for each parameter
set to optimize the scaling regime.

\begin{figure}
  \includegraphics[width=3.25in]{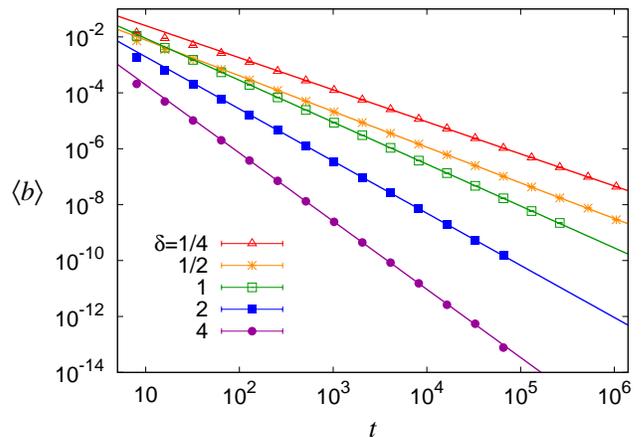}
  \caption{ Log-log plot of the average $B$ particle density versus
    time, demonstrating multiple decades of scaling for the case $p=1$
    (traps undergoing $A+A\to A$) for various diffusion constant
    ratios $\delta=D_B/D_A$.  The error bars are significantly smaller
    than the points plotted.}
  \label{fig:b_vs_t}
\end{figure}

Representative data for the $B$ particle density with $p=1$ and
varying $\delta$ values are presented in Fig.~\ref{fig:b_vs_t}, along
with the best fit power law.  Not all data points shown are used in
the fits.

We fit our data with independent errors at each time value to a power
law, choosing our minimum and maximum times according to goodness of
fit.  We estimated the uncertainty of the exponent by varying the
minimum and maximum times.  We can evaluate the effectiveness of this
procedure by comparing to two exact solutions:
\begin{itemize}
\item For $p=1$, the $B$ particle density decays like the survival
  probability in a three-walker problem \cite{Fisher88b}, giving
  \begin{equation}
    \theta = \frac{\pi}{2 \arccos [ \delta / (1+\delta) ] }.
  \end{equation}
\item For $p=0$ and $\delta=1$,
the $B$ particles behave exactly like $A$ particles, $\langle
b\rangle\sim\langle a\rangle$, giving $\theta=1/2$.
\end{itemize}
Our measured values along with their uncertainties are reported in
Table.~\ref{table:theta}.  The uncertainty estimates appear to be
reasonable.

\begin{table}
  \setlength{\tabcolsep}{10pt}
  \begin{tabular}{ccll}
    \hline\hline
    $\delta$ & $p$ & $\theta_\text{measured}$ & $\theta_\text{exact}$ \\
    \hline
    $1/4$ & $0$ & $0.4129(7)$  & \\
    $1/2$ & $0$ & $0.4434(4)$  & \\
    $1$   & $0$ & $0.5004(3)$  & $0.5$ \\
    $2$   & $0$ & $0.5899(7)$  & \\
    $4$   & $0$ & $0.7285(9)$  & \\[0.5ex]
    $1/4$ & $1$ & $1.1468(7)$  & $1.14704$  \\
    $1/2$ & $1$ & $1.2768(9)$  & $1.27607$  \\
    $1$   & $1$ & $1.4992(9)$  & $1.5$ \\
    $2$   & $1$ & $1.8650(11)$ & $1.86762$  \\
    $4$   & $1$ & $2.438(2)$   & $2.44102$ \\[0.5ex]
    $1$ & $1/4$ & $0.5923(3)$ \\
    $1$ & $1/2$ & $0.7299(10)$ \\
    $1$ & $3/4$ & $0.9581(16)$ \\
    \hline\hline
  \end{tabular}
  \caption{Measured values of $\theta$ for various diffusion constant
    ratios $\delta=D_B/D_A$ and trap reaction parameter $p$, defined
    in Eq.~(\ref{eq:trap_reaction}).  The exact values from the
    vicious walker problem are included for comparison.}
  \label{table:theta}
\end{table}

Theoretical results for $\theta$ include the exact solutions described
above, as well as Smoluchowski theory, which provides the value
$\theta_S$ given in Eq.~(\ref{eq:theta_smoluchowski}), and the RG
$\epsilon=2-d$ expansion.
Smoluchowski theory has proved to be surprisingly effective, e.g., it
correctly predicts the $A$ particle decay exponent for all dimensions
\cite{Krapivsky94}, but is an uncontrolled approximation.  By
contrast, the RG $\epsilon$ expansion is systematic, but has only been
computed to first order in $\epsilon$
\cite{Howard96,Rajesh04,VollmayrLee18}.  For completeness we provide
the result here:
\begin{equation}
  \theta =  \theta_S + \frac{1}{4}\biggl[ \frac{1+\delta}{2-p}
    + \biggl(\frac{1+\delta}{2-p}\biggr)^2 f(\delta)\biggr]\epsilon
  + O(\epsilon^2)
\end{equation}
where
\begin{equation}
  f(\delta) = 1 + 2\delta\biggl[\ln\biggl(\frac{2}{1+\delta}\biggr)
    -1\biggr]
  + (1-\delta^2)\biggl[ \text{Li}_2\biggl(\frac{\delta-1}{\delta+1}\biggr)
    -\frac{\pi^2}{6}\biggr]
  \label{eq:f}
\end{equation}
and $\text{Li}_2(v)=-\int_0^v du\, \ln(1-u)/u$ is the dilogarithm
function \cite{AbramowitzStegun}.

\begin{figure}
  \includegraphics[width=3.25in]{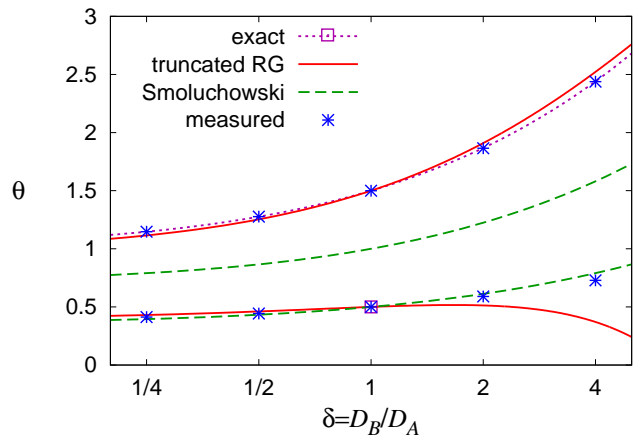}
  \caption{ Measured values of the $B$-particle decay exponent
    $\theta$ plotted versus the diffusion constant ratio, along with
    the Smoluchowski prediction, Eq.~(\ref{eq:theta_smoluchowski}),
    the RG expansion truncated at first order in $\epsilon=2-d$, and
    exact solutions.  The upper (lower) curves and points correspond
    to the $A+A\to A$ ($A+A\to 0$) trap reaction.  The error bars on
    the data are much smaller than the points plotted.}
  \label{fig:theta_p0and1}
\end{figure}

For coalescing traps, $A+A\to A$, Smoluchowski theory in $d=1$ and the
truncated $RG$ expansion with $\epsilon=1$ can be compared directly to
the vicious walker result, as was done in Ref.~\cite{Rajesh04}.  We
reproduce the comparison here as the upper curves in
Fig.~\ref{fig:theta_p0and1}, and add to the plot our measured values.
Primarily, this demonstrates that our simulations and data analysis
technique are accurate.  Also, as noted in Ref.~\cite{Rajesh04}, the
truncated RG does a remarkable job of matching the exact solution,
while the Smoluchoswki result is considerably low.

The lower set of curves and points in Fig.~\ref{fig:theta_p0and1} are
the corresponding $\theta$ values for annihilating traps, $A+A\to 0$,
where the vicious walker solution is not available.  Our measured
values for $\theta$ indicate that the Smoluchowski approximation,
while faring poorly for $p=1$, is reasonably accurate for $p=0$.  The
non-monotonicity of $\theta$ with respect to $\delta$ in the truncated
RG is likely an artifact of the truncation at $O(\epsilon)$.

\begin{figure}
  \includegraphics[width=3.25in]{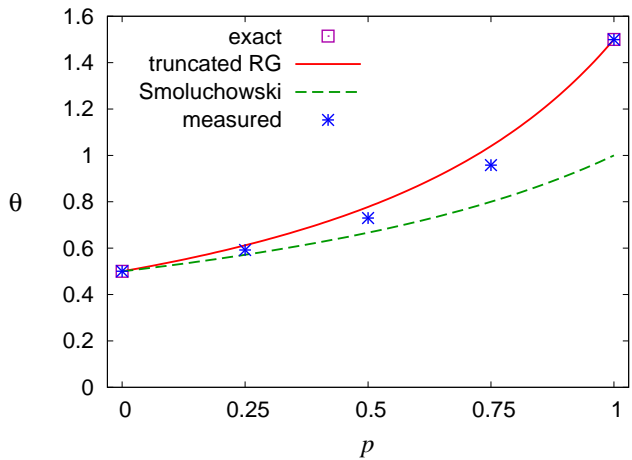}
  \caption{ A similar comparison as in
    Fig.~\ref{fig:theta_p0and1} for the equal diffusion
    constant case $\delta=1$ and varying $p$ as defined in
    Eq.~(\ref{eq:trap_reaction}).}
  \label{fig:theta_d1}
\end{figure}

Finally, in Fig.~\ref{fig:theta_d1} we present a similar comparison for
the case of equal diffusion constants but varying $p$.  Curiously, the
truncated RG expansion matches the exact solutions available at $p=0$
and $p=1$, while faring reasonably in between.

\section{Anomalous Dimension $\phi$}

\label{sec:phi}

\begin{figure}
  \includegraphics[width=3.25in]{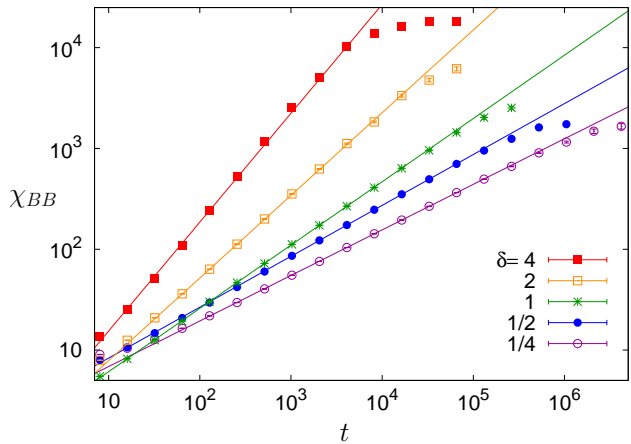}
  \caption{Log-log plot of the local fluctuations $\chi_{BB}$ plotted
    versus time, for the case $p=1$ and varying $\delta$. The straight
    lines are power law fits.  Finite-size effects are visible at
    later times, and these data are not included in the fits.}
  \label{fig:chi_BB}
\end{figure}

From the field theoretic RG calculation it was determined that $b^2$,
the square of the field associated with the $B$ density, must be
renormalized independently of the $b$ itself.  A consequence of this
renormalization is that the local fluctuations grow as a power law in
time, as measured by 
\begin{equation}
  \chi_{BB}(t)= \frac{\langle b^2\rangle - \langle b\rangle^2}{\langle
    b\rangle^2} \sim t^\phi,
  \label{eq:chi_BB}
\end{equation}
in contrast to the analogous measures
\begin{equation}
  \chi_{AA} = \frac{\langle a^2\rangle - \langle a\rangle^2}
      {\langle a\rangle^2} = -1
\end{equation}
and
\begin{equation}
  \chi_{AB} = \frac{\langle ab\rangle - \langle a\rangle \langle b\rangle}
      {\langle a\rangle\langle b\rangle} = -1
\end{equation}
which maintain constant values \cite{VollmayrLee18}.  Our measured
values for $\chi_{BB}$ versus time are plotted in
Fig.~\ref{fig:chi_BB}, for the case of coalescing traps ($p=1$).  We
observe power law behavior until the onset of finite-size effects.
Curiously, finite-size effects appear much earlier in $\chi_{BB}$ than
they do in the density, by a factor of $10^2$ or $10^3$ (compare
Fig.~\ref{fig:b_vs_t}).

\begin{figure}
  \includegraphics[width=3.25in]{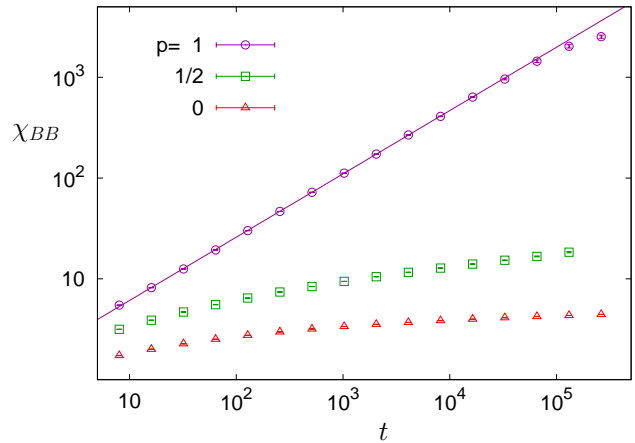}
  \caption{Log-log plot of the local fluctuations $\chi_{BB}$ plotted
    versus time, for the case equal diffusion constants $\delta=1$ and varying
    the trap reaction parameter $p$. For $p<1$ we do not reach the scaling
    regime.}
  \label{fig:chi_BB_p}
\end{figure}

We were unable to demonstrate power law behavior in $\chi_{BB}$ when the traps
are annihilating ($p=0$) or for any of the mixed reactions we simulated
($p=0.25$, $0.5$, and $0.75$), as shown in Fig.~\ref{fig:chi_BB_p}.  The data
are consistent with an asymptotic approach to a power law with a small exponent
$\phi$.

Our measured values of $\phi$ for $p=1$ are reported in
Table~\ref{table:phi}.  Our uncertainties were estimated by varying
the fitting range within the scaling regime.  For the case $\delta=1$,
an exact value of $\phi$ can be obtained by considering a four-walker
problem, where the walkers on a line are in the order $A$-$B$-$B$-$A$.  The
bracketing $A$ walkers are unaffected by any subsequent coalescence
events with exterior $A$ particles, so they may be regarded as simple
random walkers.  The $B$ particle density squared will decay as the
probability for the two interior walkers to survive until 
and meet at time $t$ \cite{VollmayrLee18}.  This exponent can be
reduced to an eigenvalue problem \cite{Helenbrook18} and the
corresponding value is reported in Table~\ref{table:phi}.

\begin{table}
  \setlength{\tabcolsep}{10pt}
  \begin{tabular}{cll}
    \hline\hline
    $\delta$ & $\phi_\text{measured}$ & $\phi_\text{exact}$ \\
    \hline
    $1/4$ & $0.452(2)$  & \\
    $1/2$ & $0.505(3)$  & \\
    $1$   & $0.628(3)$  & $0.6262475$ \\
    $2$   & $0.820(5)$  & \\
    $4$   & $1.08(4)$  & \\
    \hline\hline
  \end{tabular}
  \caption{Measured values of $\phi$ for various diffusion constant
    ratios $\delta=D_B/D_A$ and trap reaction parameter $p=1$, defined
    in Eq.~(\ref{eq:trap_reaction}).  The exact value from the
    four-walker problem is included (to 7 digits) for comparison.}
  \label{table:phi}
\end{table}

The RG calculation of $\phi$ in Ref.~\cite{VollmayrLee18} gives
\begin{equation}
  \phi = \frac{13}{24-18p}\epsilon + O(\epsilon^2),
\end{equation}
where $\epsilon=2-d$.  The truncated expansion does not compare well
quantitatively with our data, most notably in the absence of $\delta$
dependence.  Plugging in $\epsilon=1$ gives $\phi=13/6\simeq 2.17$,
which is significantly higher than the values we measured.  A
qualitative feature that the RG calculation does capture is that
$\phi$ is a strongly decreasing function of $p$.  Presumably, the RG
$\epsilon$ expansion is poorly convergent, as was found with
the simple annihilation reaction \cite{Lee94}.

\section{Correlation Functions}

\label{sec:correlations}

Associated with power law behavior with universal exponents is the
phenomenon of dynamical scaling.  These share a common origin in the
underlying RG fixed point that controls the asymptotic dynamics and
structure.  We test for this dynamical scaling by measuring the trap
and particle two-particle correlation functions, as well as their
cross-correlation function.

\begin{figure}
  \includegraphics[width=3.25in]{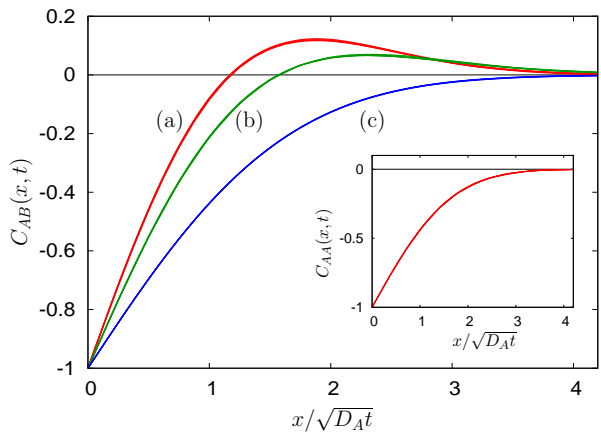}
  \caption{ Scaling collapse of the measured correlation functions for
    times ranging over three decades.  The cross correlation function
    $C_{AB}(x,t)$ parameters are (a) $p=1$,
    $\delta=1/4$, (b) $p=1$, $\delta=1$, and (c) $p=0$, $\delta=1$.
    The inset shows the measured $C_{AA}(x,t)$ for $p=0$,
    $1/2$, and $1$, as well as the exact solution,
    Eq.~(\ref{eq:C_AAsolution}), with striking agreement.}
  \label{fig:corr_ab_aa}
\end{figure}

We first consider the traps, which undergo the single-species $A+A\to
0, A$ reactions.  An exact solution for the correlation function in
$d=1$ was obtained by Masser and ben-Avraham, with the result
\cite{Masser01}
\begin{equation}
  C_{AA}(x,t) = \frac{\langle a(x,t) a(0,t)\rangle -
    \langle a(t)\rangle^2}{\langle a(t)\rangle^2} \sim f_{AA}(x/\sqrt{D_At})
\end{equation}
where
\begin{equation}
  f_{AA}(z) = -e^{-z^2/4} + \sqrt{\frac{\pi}{8}} z
  e^{-z^2/8}\,\textrm{erfc}(z/\sqrt 8).
  \label{eq:C_AAsolution}
\end{equation}
Interestingly, this result applies to both annihilating and coalescing
particles, as well as mixed reactions.  We measured these correlation
functions via the Monte Carlo realization of our trap dynamics and
found convincing scaling collapse and perfect agreement with the exact
solution, as shown in the inset of Fig.~\ref{fig:corr_ab_aa}.

We next turn to the cross correlation function
\begin{equation}
  C_{AB}(x,t) = \frac{\langle a(x,t) b(0,t)\rangle
    - \langle a(t)\rangle\langle b(t)\rangle}
         {\langle a(t)\rangle\langle b(t)\rangle},
\end{equation}
which is plotted in Fig.~\ref{fig:corr_ab_aa}. With our hybrid
simulation method we measure the correlation between the realized $A$
particles and the associated $B$ probability distribution.  The data
again exhibit convincing scaling collapse, with a scaling function
that depends on the parameters $\delta$ and $p$.  Both $C_{AA}$ and
$C_{AB}$ exhibit anti-correlations at short distances, a direct
consequence of the $A+A\to (0,A)$ and $A+B\to A$ reactions.  However,
depending on the parameter values, the cross-correlation function
$C_{AB}$ can be non-monotonic with positive correlations at larger
separation.  We depict three choices of parameters in
Fig.~\ref{fig:corr_ab_aa}, but we found similar scaling collapse for
all investigated cases.

Finally, we turn to the $B$ particle correlation function defined in
Eq.~(\ref{eq:BBcorrelations}) and measured by the sampled set of
$B$ particle distributions.  Since the $B$ particles do not react
with each other, we do not expect them to be anti-correlated at short
distances.  Instead, a surviving $B$ particle is likely to be found in
a region with few $A$ traps nearby, which results in an enhanced
probability of other $B$ particles nearby, i.e.,\ positive
correlations.

\begin{figure}
  \includegraphics[width=3.25in]{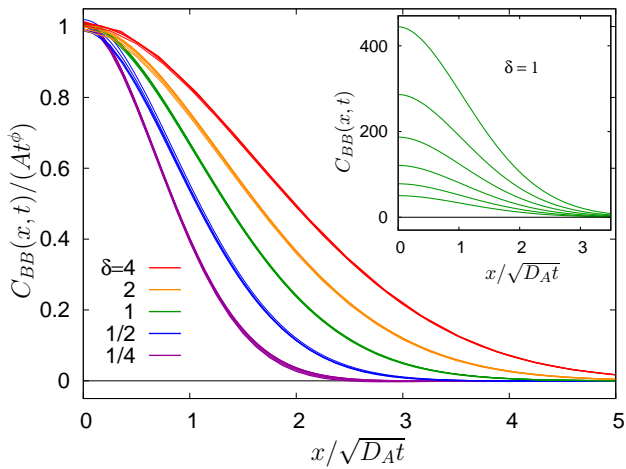}
  \caption{ Scaling collapse of the measured correlation functions
    $C_{BB}(x,t)$ for times ranging over two decades, which requires
    rescaling the vertical axis by $\chi_{BB} \sim At^\phi$.  All
    plots are for coalescing traps ($p=1$).  The inset shows $C_{BB}$
    for $\delta=1$ without the vertical rescaling; the intercept is
    increasing with time.}
  \label{fig:corr_bb}
\end{figure}

Our measured values for correlation function confirm this, as shown in
Fig.~\ref{fig:corr_bb}.  The inset shows that when $C_{BB}(x,t)$ is
plotted versus the scaled distance $x/\sqrt{D_At}$, as was done in
Fig.~\ref{fig:corr_ab_aa}, we do not find collapse, but rather the
correlations are growing in magnitude with time.  However, when we
also scale the vertical axis by the expected $\chi_{BB}\sim At^\phi$,
with $A$ and $\phi$ taken from our fitted values, we indeed see
scaling collapse, as shown in the main part of Fig.~\ref{fig:corr_bb}.
Thus we have confirmed the RG prediction of the scaling form in
Eq.~(\ref{eq:BBcorrelations}).

The scaled correlations for $p=1$ show a significant dependence on the
diffusion constant ratio.  The similarity of the scaling functions
suggest that a rescaling of the horizontal axis to the form
$x/\sqrt{D_A^{1-k}D_B^kt}$ might collapse all measured functions to a
single curve.  Indeed, the value $k=0.60$ comes close though slight
differences are observable.  Evidently the power-law dependence
captures a dominant feature of the $\delta$-dependence on the scaling
function, but is not an exact result and there is currently no
theoretical basis to expect such behavior.

When $p<1$ we cannot make a scaling plot similar to
Fig.~\ref{fig:corr_bb} since we are unable to simulate late enough to
get into the regime where $\chi_{BB}$ is a power law.  If we instead
rescale the vertical scale by $C_{BB}(0,t)$ we find reasonable scaling
collapse, suggesting the shape of the correlation function converges
more quickly than $\chi_{BB}$ itself.

\section{Summary}

\label{sec:summary}

We have developed a hybrid simulation method for the coupled
two-species reactions $A+B\to A$ and $A+A\to(0,A)$ that involves a
Monte Carlo simulation of the traps combined with the full probability
distribution for the particles.  This method provides significant
improvement for statistics and avoids the problem of vanishing $B$
particle numbers.

With this technique, we explored the behavior of this
reaction-diffusion system for a variety of diffusion constant ratios
and trap reaction types.  In all cases we were able to obtain
convincing power law decay of the $B$ particle density and measure the
decay exponent to $0.1\%$ accuracy, as shown in
Table~\ref{table:theta}, with results that are consistent with known
exact values.  Our data were compared with theoretical results from
the RG $\epsilon=2-d$ expansion and from Smoluchowski theory.

We further tested the recently calculated anamolous dimension in the
$B$ particle correlation function, or equivalently in the local
fluctuations of the $B$ particles: $\chi_{BB} = C_{BB}(0,t)\sim t^\phi$.
For the case of coalescing traps we were able to obtain multiple decades
of power law scaling and measure the exponent $\phi$ to $0.5\%$
accuracy (see Table~\ref{table:phi}).  Our measured values do not
match the truncated RG calculation, but are consistent with one exact
value.

We have also tested for universality by varying the trapping reaction
probability $p'$, defined in Eq.~(\ref{eq:b_reaction_update}).  We
confirmed that the exponents $\theta$ and $\phi$ and the correlation
functions are not dependent on this parameter, consistent with them
being universal functions of $\delta$ and $p$.  In contrast, the
amplitude of the density decay $\langle b\rangle \sim A t^{-\theta}$
does dependent on $p'$ and is nonuniversal.

It is noteworthy that the power law behavior in the correlation
function $C_{BB}(x,t)$ and fluctuations $\chi_{BB}$ encountered
finite-size effects much earlier than the density $\langle b\rangle$.
From Fig.~\ref{fig:chi_BB} we see finite size effects entering around
$t=3\times 10^4$ for the equal diffusion constant case, at which time
the diffusion length is $\sqrt{Dt} \sim 100$ in a system of size
$3\times 10^7$.  The origin of this extreme sensitivity merits further
investigation, both analytically and numerically.

\acknowledgments

R.~C.~R. was supported by NSF Grant No. REU-0097424.  B. P. V.-L.
acknowledges the hospitality of the University of G\"ottingen, where
this manuscript was completed.


\end{document}